\documentclass[prl,twocolumn,superscriptaddress,a4paper]{revtex4}
\usepackage{graphicx}
\usepackage{epstopdf}
\usepackage{amssymb,amsmath}
\usepackage{epsf}

\newcommand{\ket}[1]{|#1\rangle}
\newcommand{\braket}[1]{\langle#1\rangle}

\usepackage{amssymb}
\usepackage{amsmath}

\begin{document}

\title{Ramsey interference in one dimensional systems: 
The full distribution function of  fringe contrast as a probe 
of many-body dynamics}

\author{Takuya Kitagawa}
\affiliation{Physics Department, Harvard University, Cambridge,
MA 02138, USA}
\author{Susanne Pielawa}
\affiliation{Physics Department, Harvard University, Cambridge,
MA 02138, USA}
\author{Adilet Imambekov}
\affiliation{Department of Physics and Astronomy, Rice University, Houston, Texas 77005, USA}
\author{J\"{o}rg Schmiedmayer}
\affiliation{Atominstitut,TU-Wien,Stadionallee 2, 1020 Vienna, Austria}
\author{Vladimir Gritsev}
\affiliation{Physics Department, University of Fribourg, Chemin du Musee 3, 1700 Fribourg, Switzerland}
\author{Eugene Demler}
\affiliation{Physics Department, Harvard University, Cambridge,
MA 02138, USA}

\begin{abstract}
We theoretically analyze  Ramsey interference experiments in one dimensional
quasi-condensates and obtain explicit
expressions for the time evolution of full distribution functions of fringe contrast. We show 
that distribution functions contain unique signatures of the many-body mechanism of decoherence.
We argue that Ramsey interference experiments provide a powerful tool for analyzing strongly 
correlated nature of 1D interacting systems. 
\end{abstract}

\date{\today}

\maketitle

{\it Introduction} 
Recent progress in the field of ultracold  atoms not only expanded
our understanding of equilibrium properties of interacting 1d Bose gases 
\cite{hofferberth,bloch} but posed new theoretical challenges by 
studying far-from-equilbrium dynamics of such systems. 
Recent experiments addressed such questions as thermalization 
and integrability\cite{weiss}, decoherence after the splitting of two condensates\cite{hofferberth2} and 
spin dynamics of two component Bose mixtures\cite{widera}. Motivation for such experiments comes not only from
the basic interests in many-body dynamics\cite{shlyapnikov}
but also from possible applications of ultracold atoms such as quantum information processing \cite{zoller} and 
interferometric sensing\cite{review_pritchard}.
In this paper we theoretically analyze the decoherence dynamics of Ramsey interference fringes 
in one dimensional quasi-condensates. 
Such systems have been considered for possible applications in atomic clocks and 
quantum enhanced metrology 
\cite{sorensen, kitagawa}. In this paper we show that Ramsey interferometer is also a 
powerful tool for studying many-body dynamics of low dimensional quantum systems.
We find that decoherence of Ramsey fringes
is strongly affected by the multimode character of one dimensional systems. 
Moreover we will
demonstrate that time evolution of the full distribution function of fringe contrast 
provides unique signatures of this many-body decoherence mechanism \cite{pnas}. 
The idea of using noise and distribution functions to characterize equilibrium many-body states
of ultracold atoms has been discussed in several theoretical papers
\cite{altman2}
and applied 
in experiments\cite{dalibard,hofferberth}. However there has been so far no application of this approach to 
non-equilibrium dynamics. This paper constitutes the first proposal 
to study non-equilibrium dynamics of ultracold atoms with quantum noise.

The role of interactions in Ramsey 
interferometers with BEC was first addressed in the pioneering 
paper of Kitagawa and Ueda\cite{kitagawa}. They 
 used single mode approximation to 
 predict the interaction induced decoherence of 
Ramsey fringes along with  the appearance of 
spin squeezed states. 
Their work stimulated ideas for quantum-enhanced 
metrology that take advantage of spin squeezed states formed in 
interacting BECs \cite{sorensen,amrey}.
For the analysis of one dimensional quasi-condensates, however, the single mode approximation 
cannot be applied because these systems do not have 
macroscopic occupation of a single state even at zero temperature.
The non mean-field character of the multi-mode spin dynamics in 1D 
quasi-condensates was first reported in the experiments of
Widera et al \cite{widera}. However, this work did not provide the definitive 
demonstration of the many-body origin of decay. 
In the following, we argue that
unambiguous signatures  of the multimode decoherence 
are found in the full distribution function of the Ramsey fringe amplitudes. 
Such distribution functions should be accessible in experiments with 
1D quasi-condensates realized on Atom Chips\cite{atomchip}, 
because such systems do not 
average over multiple tubes and thus allow the measurements of shot-to-shot 
fluctuations \cite{hofferberth}.

Now we describe the Ramsey sequence considered in this paper. Here we 
identify two hyperfine states as spin up and down states.
Ramsey sequence is carried out as follows: 
(i) all spins of the atoms are prepared in the spin up state; 
(ii) $\pi/2$ pulse is applied to rotate each spin into the $x$ direction; 
(iii) spins freely evolve (precess) for time $t$; 
(iv) another $\pi/2$ pulse is applied to map the
transverse spin component to the $z$ direction, which is then measured.
Measurements yield a net spin $\vec{S}_l$ for a 
segment of length $l$ and we assume that $l$ is smaller than the the system size
but large enough to contain large number of particles $N_l>>1$. 
In such a case, the simultaneous measurements of 
$S_l^x$ and $S_l^y$ are possible, because 
even though operators $S_l^x$ and $S_l^y$ generally do not commute, non-commutativity gives 
only corrections of the order of $1/\sqrt{N_{l}}$ 
relative to the average values \cite{Polkovnikov_europhysletters}. 
Commutativity of $S_l^x$ and $S_l^y$ implies, in particular, that we can define the
joint distribution function $P_l^{x,y}$ for the two transverse spin components
$S^x_{l}$and $S^y_{l}$. In experiments, measurements of $P_l^{x,y}$ is 
possible by mapping the spin orientations in $x-y$ plane to $z$ direction
by $\pi/2$ pulse, followed by local measurements of $S_{z}$ \cite{ketterle}.  
$S^{x}_{l}$ and $S^{y}_{l}$ as well as the magnitude of spin $S^{\perp}_l = \sqrt{(S_l^{x})^2+ (S_l^{y})^2}$
can be found by taking the integration over $l$. 
The analytic solution for the time evolution of 
$P_{l}^{x,y}$ constitutes the main result of this paper.
In addition, we  assume that $l$ is larger than the spin
healing length $\xi_s$, so we can use Tomonaga-Luttinger liquid approach
to describe the collective spin dynamics (see also below). 
For simplicity we work in the rotating frame of the Larmor precession, and consider 
the spins before the last $\pi/2$ pulse. 
Then the amplitude of Ramsey fringes, as it is conventionally defined,
corresponds to  $S^x_{l}$. 

\begin{figure}[t!]
\begin{center}
\includegraphics[width = 8.5cm]{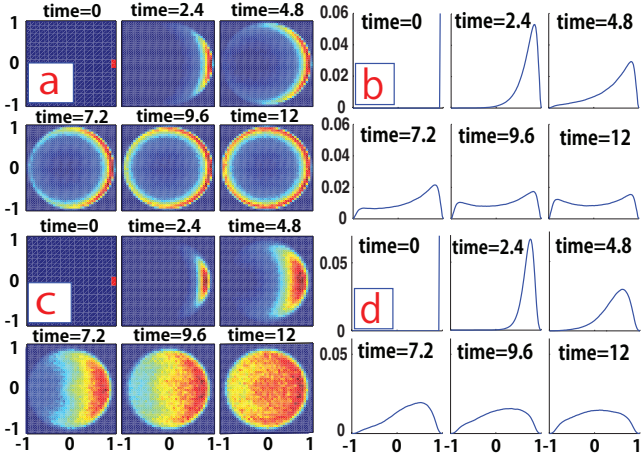}
\caption{ (a),(c): Evolution of joint FDF $P^{x,y}_{l}$
with short integration length $l/\xi_{s}=10$(top, (a)) and long integration length
$l/\xi_{s}=30$(bottom, (c)). 
(b),(d): Corresponding FDF for spin $x$,
$P^{x}_{l}$ with short integration length $l/\xi_{s}=10$
(top, b)) and long integration length $l/\xi_{s}=10$(bottom, d)). 
Here $L/\xi_{s} =200,K_{s} = 20$. 
Time is measured in units of $\xi_{s}/c_{s}$.}
\label{jointfigure}
\end{center}
\end{figure}

Our main results are summarized in Fig. \ref{jointfigure} and can be understood from the following physical arguments. 
Strong fluctuations present in 1D systems forbid the existence of long range coherence\cite{giamarchi}, 
and spatial fluctuations coming from different wavelength strongly affect the dynamics in 1D. 
 Among those, fluctuations with wavelengths longer than the integration length
$l$ rotate $\vec{S}_l$ as a whole. So they decrease $S^x_l$ but not the magnitude of the spin 
$S^{\perp}_l = \sqrt{(S_l^{x})^2+ (S_l^{y})^2}$. Fluctuations
with wavelengths shorter than $l$ decrease both $S_l^x$ and $S^{\perp}_l$ simultaneously. 
Fig.\ref{jointfigure} (a) (b) show the situation where fluctuations with wavelength larger than $l$
dominate the dynamics.
In (a), we see that
the magnitude of the spin  $S^{\perp}_l$ decays only slightly from the initial state  
but the direction of the spin is randomized during the time evolution. 
Note that in this case 
the distribution function of $S_l^x$ has a very peculiar
shape with two peaks at large positive and negative values. We call this regime, "spin diffusion" regime. 
Fig.\ref{jointfigure}  (c) (d) show the situation where fluctuations with wavelengths shorter than $l$ dominate. 
In this case $S^{\perp}_l$ and $S_l^x$ decay in the same timescale. We call this regime, "spin decay" regime.
Below we argue that the crucial parameter of the system is a dimensionless ratio proportional to the
length of the integration region $l_0=\frac{\pi^2 l}{4K_{s} \xi_{s}}$. 
When $l_0 \le 1$ the system is in the spin diffusion regime and the other limit $l_0>>1$ is the spin decay regime.

{\it Model}. Following the first $\pi/2$ pulse we have a two component Bose mixture with equal densities of
both species. Tomonaga-Luttinger liquid (TLL) approach, which we use in this paper,
focuses on the linearly dispersing modes in the low energy part of the spectrum. 
For simplicity we consider the case when interaction parameters satisfy
$g_{\downarrow\downarrow}= g_{\uparrow\uparrow}$ \cite{future_paper}. This condition can be 
reached for the hyperfine states $\ket{F=1,m_F=-1}$ and $\ket{F=2,m_F=+1}$ 
of $^{87}$Rb that are commonly used in experiments
\cite{widera,hofferberth}. 
When this is the case, the charge and spin parts of the Tomonaga-Luttinger Hamiltonian decouple
and the spin part of the Hamiltonian is given by
\begin{eqnarray}
H_{s} & =& \frac{c_{s}}{2} \int^{L/2}_{-L/2}dr\left[ \frac{K_{s}}{\pi} 
\left(\nabla \phi_{s}(r)\right)^2 +\frac{\pi}{K_{s}} n^2_{s}(r) \right]
 \label{hamiltonian}\\
& =&  \sum_{k \neq 0} c_{s}|k| b_{s,k}^{\dagger}b_{s,k}  + \frac{c_{s} \pi}{2 K_{s}} n_{s,0}^2,
\label{harmonic}
\end{eqnarray}
Here $L$ is the total system size, 
$n_{s}(r)$ describes the local spin imbalance (i.e. $z$ component of the spin) $n_{s} =
\psi^\dagger_\alpha (\frac{1}{2} \sigma^z_{\alpha\beta}) \psi_\beta $ and $n_{s,k}$ is the 
Fourier transform of $n_{s}(r)$. 
$\phi_{s}(r,t)$ describes the direction of the transverse spin component
$\rho e^{i\phi_{s}} = \psi^\dagger_\alpha (\frac{1}{2}\sigma^+_{\alpha\beta}) \psi_\beta
$ with $\rho$ being the average density for each species. Variables $n_{s}$ and $\phi_{s}$ obey canonical commutation relations
$[n_{s}(r),\phi_{s}(r')]= -i \delta(r-r')$. $K_{s}$ is spin Luttinger parameter
representing the strength of interactions\cite{widera}, and $c_{s}$ is spin-wave velocity.
Hamiltonian (\ref{harmonic}) has momentum cutoff set by the inverse of the spin 
healing length $\xi_s^{-1}$. In the weak interaction limit, these parameters are related to 
physical parameters as $c_{s} = \sqrt{ g_{s} \rho/ m}$, $\xi_{s} = \frac{\pi c_{s}}{ g_{s} \rho}$ and
$K_{s} = \xi_{s} \rho/2$, where $m$ is the mass of the particle and $g_{s} = \frac{4 \pi^2 \hbar^2}{m}
\frac{a_{\uparrow \uparrow} + a_{\downarrow \downarrow} -2 a_{\uparrow \downarrow} }{2}$ is the interaction 
strength in spin channel. In this paper, we focus on the regime $g_{s}>0$ when the system is 
miscible. Our approach can be extended to $g_{s}<0$ but will be limited to times before $z$ magnetization
per atom becomes of the order of one. 
The first and second term of (\ref{harmonic}) correspond to $k \neq 0$ and $k=0$ part
of the Hamiltonian, respectively. Here operators $b^\dagger_{s,k}$ 
create spin excitations with momentum $k$, and these spin excitations are the main
focus of our study. 

Transverse part of the spin operator $\vec{S}_l$ 
is given by
\begin{eqnarray}
S^x_{l} = \int_{-l/2}^{l/2} dr \rho \cos(\phi_{s}(r)),  \textrm{ }S^y_{l} = \int_{-l/2}^{l/2} dr \rho \sin(\phi_{s}(r)), \label{sdefinition} 
\end{eqnarray}
When describing spin dynamics one typically
considers time evolution of the expectation values
$\langle S^a_{l}(t) \rangle$.  However
important information is also contained in the shot to shot 
fluctuations of  $S^a_{l}(t)$. 
Such quantum noise is captured by
full distribution functions (FDF) of spin operators,$P^{a}_{l}(\alpha,t) $
\cite{pnas}. 
In particular, high moments of 
$S^a_{l}(t)$ can be obtained from these FDFs $P^{a}_{l}(\alpha,t) $. 
Physically $P^{a}_{l}(\alpha,t) d\alpha$ is the probability that a single
measurement of the spin operator $S^a_{l}$ 
at time $t$ gives the value between $\alpha$ and $\alpha+d\alpha$. 
In the experiments, $P^{a}_{l}(\alpha,t) $ can be obtained by 
making histograms of the measurement results of $S^a_l(t)$.

To describe time evolution of spin operators (\ref{sdefinition})
we need to characterize the initial state of the system after the first
$\pi/2$ rotation. The difficult part is translating the initial state of the microscopic
language, where spins of all atoms pointing in the $x$ direction, to 
the one in terms of the coarse grained degrees of freedom
$\phi_s(r)$ and $n_s(r)$. Classically one expects the initial state to be the eigenstate
of $\phi_s(r)$ with eigenvalue zero for all $r$. However such 
state is unphysical in quantum mechanics since it leads 
to infinite uncertainty of the conjugate variable $n_s(r)$, and thus, to infinite energy.
More sensible initial state is a squeezed state of harmonic oscillator 
(\ref{hamiltonian}), which has reduced uncertainty in $\phi_s$ at the expense
of enhanced fluctuations in $n_s$\cite{average,widera}. 
To determine parameters of this state
we observe that spins of individual atoms are independently rotated by the first $\pi/2$ pulse into 
$x$ direction, so the initial state satisfies 
$\langle  S^{z}(r) S^{z}(r') \rangle = \frac{\rho }{2} \delta(r-r')$. 
Thus we find a Gaussian state for the spin operator $S^{z}$ in momentum space 
with fluctuations $\rho/2$ for all $k$:
\begin{equation}
| {\psi_{0}}  \rangle = \frac{1}{\mathcal{N}} \exp{\left(\sum_{k\neq 0} W_{k} b_{s,k}^{\dagger} b^{\dagger}_{s,-k} \right)} | {0} \rangle
| {\psi_{s,k=0}} \rangle,
\label{psi_0}
\end{equation} 
where $2W_{k} =  {(1-\alpha_{k})}/{(1+\alpha_{k})}$, 
$\alpha_{k} = {|k|K_{s}}/{\pi \rho}$ and 
$\mathcal{N}$ is the overall normalization of the state. For the uniform part of the
spin operator we also have a squeezed state
$ \langle n_{s,0} | {\psi_{s,k=0}} \rangle = \exp(-1/(2\rho) n_{s,0}^2) $.
We note that model (\ref{hamiltonian}) 
has a short distance cut-off so 
the $\delta$ function in $\langle  S^{z}(r) S^{z}(r') \rangle $ should be understood as 
rounded off on the scale of  $\xi_s$, which is implicit in the momentum cut-off in 
Eq. (\ref{psi_0}). 

Time evolution of the state  (\ref{psi_0})  leads to $W_k \rightarrow W_k e^{2ic_s|k|t}$. From the resulting 
expression for the state at time $t$, one can readily calculate the decay of Ramsey fringes given by
$\braket{S^{x}_{l}(t)}$, which is independent of integration length $l$ (See also \cite{widera, average}).  
To calculate time evolution of FDF, we define instantaneous annihilation operators $\gamma_{ks}(t)$ such that application 
of $\gamma_{ks}(t)$ on the state $\exp{\left( W_{k} e^{2ic_s|k|t } b_{s,k}^{\dagger} b^{\dagger}_{s,-k} \right)} | {0} \rangle$
gives zero. Using operators  $\gamma_{ks}(t)$, one can apply the approach described 
in Refs. \cite{hofferberth,pnas}
for calculating distribution functions of equilibrium systems. After direct calculation we find
 \cite{future_paper}
 \begin{eqnarray}
P^{x,y}_{l}(\alpha,t) &=&  \prod_{k} \int^{\infty}_{-\infty} e^{-\lambda_{rsk}^2/2} d\lambda_{rsk} 
 \int^{\pi}_{-\pi} d\lambda_{\theta sk}\nonumber \\
&& \delta\left(\alpha - \rho \int^{l/2}_{-l/2} dr e^{i\chi( r,t, \{\lambda_{jsk} \})} \right),
\label{jointdist}\\
\chi(r, t,\{\lambda_{jsk} \}) & =&
\sum_{k} \lambda_{rsk} \sqrt{\frac{\braket{|\phi_{s,k}(t)|^2}}{L} } \sin(kr+\lambda_{\theta sk}), 
  \nonumber 
  \end{eqnarray}
\begin{eqnarray}
\braket{|\phi_{s,k \neq 0}|^2} &=& 
\left(\frac{\pi \rho}{|k| K_{s}} \right)^2 \frac{\sin^2(c_{s}|k|t) }{2\rho}
   + \frac{\cos^2(c_{s}|k|t) }{2\rho},                \nonumber \\
   \braket{|\phi_{s,k = 0}|^2} & = &  \frac{1}{2\rho} + \left(\frac{c_{s}\pi t}{K_{s}}\right)^2 \frac{ \rho}{2},
   \label{phase_fluctuation}
\end{eqnarray}  
where the real and imaginary part of $\alpha$ corresponds to $x$ and $y$ component of 
$\vec{S}_{l}$, respectively. 
Eq. (\ref{jointdist}), (\ref{phase_fluctuation}) allow a simple physical interpretation. Function
$\chi(r,t,\{\lambda\})$ defines the local direction of transverse magnetization, which results
from the summation over spin-wave like modes 
$\sin(kr+\lambda_{\theta sk})$. Amplitudes of individual modes are given 
by the time dependent expectation values $\langle |\phi_{s,k}(t)|^2\rangle$ and by the
set of random variables $\lambda_{rsk}$ drawn from a Gaussian ensemble.
Eq. (\ref{jointdist}), (\ref{phase_fluctuation}) reflect the key feature of dynamics of the quadratic Luttinger model  (1): 
initial  Gaussian state for $\phi_{s,k}$ remains Gaussian at all times \cite{endnote_g}. 

Time evolution of $\langle |\phi_{s,k}(t)|^2\rangle$ following the first $\pi/2$ rotation can be understood as free dynamics
of a harmonic oscillator. From the conjugate nature of $\phi_{s,k}$ and $n_{s,k}$
we find $\langle |\phi_{s,k}(0)|^2\rangle = \frac{1}{4} \frac{1}{\braket{|n_{s,k}(0)|^2}} = \frac{1}{2\rho}$ at $t=0$. 
Subsequently $\langle |\phi_{s,k}(t)|^2\rangle$
oscillates between the minimal value in the 
initial state and some maximum value $\braket{|\phi_{s,k}|^2}_{max}$ at the frequency of a harmonic 
oscillator $c_{s}|k|$. $\braket{|\phi_{s,k}|^2}_{max}$ can be estimated 
from energy conservation. Since the initial state was squeezed with respect to $\phi_{sk}$,
most of the energy of the mode is stored in the interaction term  $|n_{s,k}|^2$. 
 Therefore the total energy of the harmonic oscillator for momentum $k$ can be approximated by
 $\frac{\pi c_{s} \rho}{K_{s} }$, which in turn gives
 $\braket{|\phi_{s,k}|^2}_{max} \sim \frac{2\pi^2 \rho}{K^2_{s} k^2 } = \frac{1}{2\rho} \left(\frac{\pi \rho}{|k| K_{s}} \right)^2$.
 These considerations lead to the dynamics of phase fluctuation amplitude of the form in (\ref{phase_fluctuation}).  
 We note that the spin fluctuations are dominated by small momentum
 since the maximum fluctuation amplitude is suppressed as $1/k^2$ for large momentum. 
 This justifies our analysis based on the Tomonaga-Luttinger theory.

   \begin{figure}[t!]
\begin{center}
\includegraphics[width = 8.5cm]{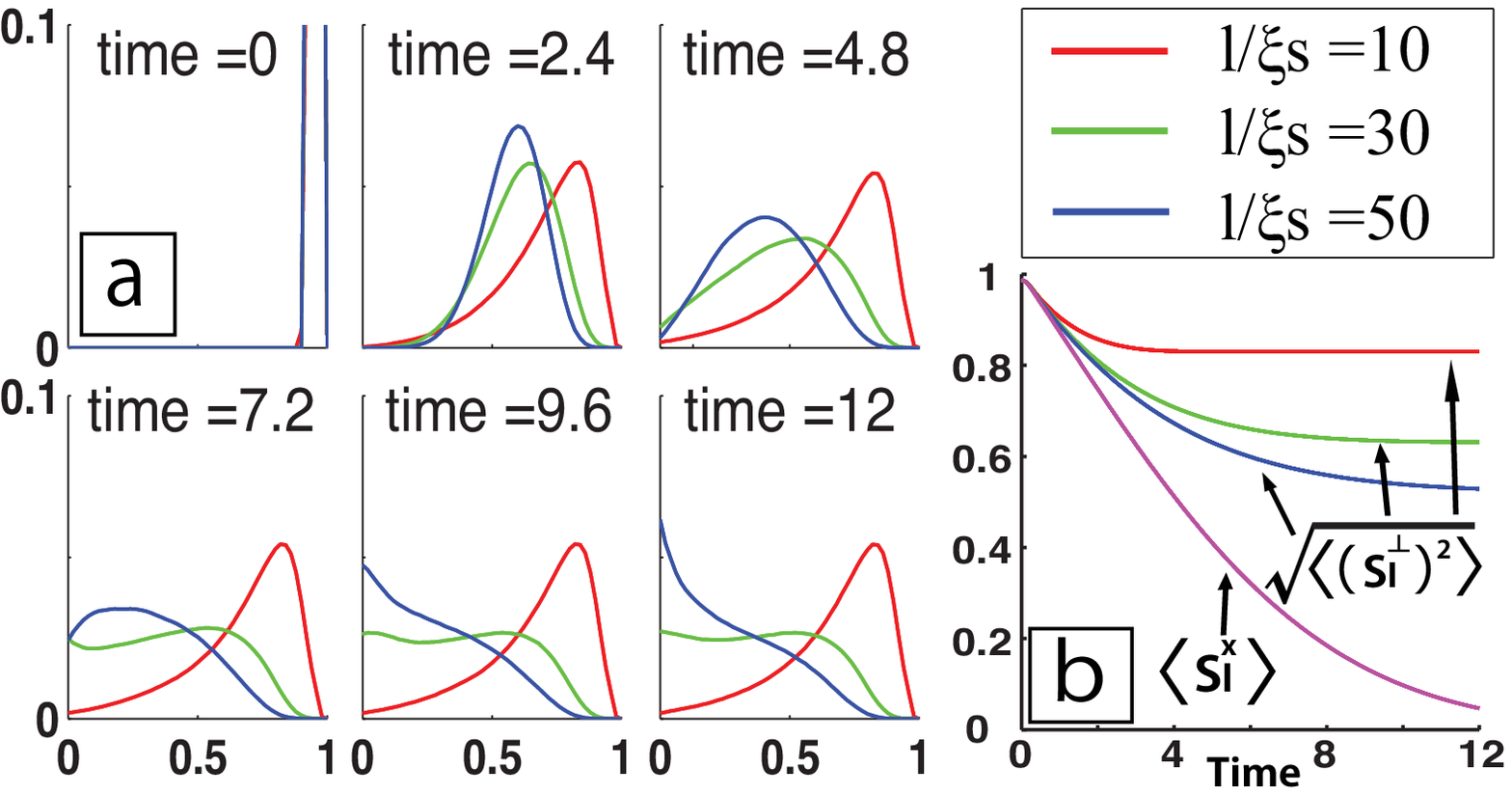}
\caption{(a):Time evolution of the distribution $P^{\perp}_{l}$ 
for the magnitude of spin $\left(S^{\perp}_{l} \right)^2$.
(b): time evolution of $\braket{S_{l}^{x}}$ and $\sqrt{\braket{\left(S^{\perp}_{l} \right)^2}}$ 
with various integration length $l/\xi_{s} = 10,30,50$.
Here we set $K_{s} =20$, $L/\xi_{s} = 200$.  }
\label{timescale}
\end{center}
\end{figure}
 
Results of numerical plots based on Eq. (\ref{jointdist}), (\ref{phase_fluctuation})  
are shown in Fig. \ref{jointfigure}. In Fig. \ref{timescale} (a) we also present
the distribution function $P^{\perp}_{l}$ of the magnitude square of the integrated  spin $\left(S^{\perp}_{l} \right)^2$, which
clearly demonstrates the difference between the "spin diffusion" and "spin decay" regimes. 
The character of these  distribution functions can be understood from 
the following arguments. We first discuss the "spin diffusion" regime, where the characteristic wavelength
of spin fluctuations is longer than the integration length $l$ (Fig. \ref{jointfigure} (a), (b)). 
In this regime, all spins within $l$
essentially point in the  same direction  and $S^{\perp}_{l}$ remains large even after a long time evolution. 
Thus we find a peak at $\left(S^{\perp}_{l} \right)^2 \approx 1$
in the distribution function $P^{\perp}_{l}$(red line).
In the other regime of "spin decay" (Fig. \ref{jointfigure} (c), (d)), the typical length scale of spin fluctuations
is shorter than the integration length $l$. In this case integration of spins over $l$ is akin to taking a 
random walk in 2D plane and accordingly, distribution for $\left(S^{\perp}_{l} \right)^2$ approaches 
exponential form with a peak at $\left(S^{\perp}_{l} \right)^2 =0$ in $P^{\perp}_{l}$(blue line). 
 In the intermediate regime, we observe two peak structure for $P^{\perp}_{l}$, 
 where the distribution exhibits  both characteristic peaks(green line). 
 We can understand the condition that separates these "spin decay" and "spin diffusion" type dynamics
 in the following way. 
Deviation of spin angles at $r=l$ relative to $r=0$ can be estimated from
 $\Delta \chi \approx  \frac{1}{\sqrt{L}} \sum_{k} \lambda_{rsk} \sqrt{\braket{|\phi_{s,k}|^2}_{max}} \sin(kl)$. 
 A typical magnitude of $\Delta \chi$ is given by 
$\braket{ (\Delta \chi)^2}$ where the average is taken over fluctuations of $\lambda_{rsk}$. 
The factor $\sin(kl)$ in $\Delta \chi$ effectively limits momentum integration range 
to $k>2\pi/l$, so $\braket{ (\Delta \chi)^2} \approx \frac{\pi^2 l }{2 K_{s} \xi_{s}}$. 
When $\braket{ (\Delta \chi)^2}^{1/2}$ is smaller than $2\pi$ the system is in the "spin diffusion" regime.
When  $\braket{ (\Delta \chi)^2}^{1/2}$ becomes of the order of $2\pi$ and larger, the system enters
the "spin decay" regime. The crossover takes place  around $ \frac{\pi^2 l}{4K_{s} \xi_{s}} \sim 1$.
 
 In the "spin diffusion" regime, the dynamics of $S^{\perp}_{l}$ and $S_{l}^{x}$ display different 
 time scales as can be seen in Fig. \ref{timescale} (b). 
 In order to understand
 this separation of time scale, we note that the magnitude of
 the integrated spin $S^{\perp}_{l}$ is only affected
 by fluctuations with short wavelengths, $\lambda <l$, for which dynamics takes place at short time scale. 
 Hence $\braket{S^{\perp}_{l}}$ 
 decays until the time $t_{\perp} \approx 2\pi l/c_{s}$(see Eq. (\ref{phase_fluctuation})) and then it reaches a saturated value. 
 On the other hand $S^{x}_{l}$ is affected by excitations of all wavelength,
 so $\braket{S^{x}_{l}}$ decays until it reaches $0$. These behaviors are shown in 
 Fig. \ref{timescale} (b), where we compared the decay of $\braket{S^{\perp}_{l}}$ for various segment
 length $l$ and $\braket{S^{x}_{l}}$. 
 
  Non-trivial time evolution of the distribution functions $P_{l}^{x}$, $P^{x,y}_{l}$ and especially 
 the striking contrast of the "spin diffusion" and "spin decay" regimes should provide unique signatures of the non-mean-field character and multimode dynamics of 1D systems. 
 
 Before concluding this paper we point out that our analysis can be extended to the problem of splitting
 a single 1D quasi-condensate into two, as it was done with an RF potential in experiments
 reported in Refs.\cite{hofferberth2}. While earlier theoretical work focused on the time decay of the
 average fringe contrast\cite{average}, the method developed in this paper can be used to study the time evolution of the full distribution function.

{\it Summary}.
We provided theoretical analysis of Ramsey interference experiments with one dimensional
quasi-condensates. We discussed time evolution of the full distribution functions of fringe contrast
and showed that  they contain unique signatures of the many-body dynamics
of one dimensional systems. 
This work was supported by the
NSF grant DMR-0705472, Harvard MIT CUA, DARPA OLE program, AFOSR MURI, and Swiss NSF.


\begin{thebibliography}{99}
\bibitem{endnote_g}
Temperature determines the density fluctuations before the $\pi/2$ spin rotation. However, 
spin and density fluctuations are decoupled in the system we consider so that our 
results do not depend on temperature. Corrections due to small $g_{\uparrow \uparrow} \neq
g_{\downarrow \downarrow}$ introduce temperature dependence, but they do not lead to 
qualitative changes \cite{future_paper}.
\end{thebibliography}
\end{document}